\documentstyle[aps,psfig,twocolumn]{revtex}
\begin{document}
\draft
\flushbottom
\twocolumn[
\hsize\textwidth\columnwidth\hsize\csname @twocolumnfalse\endcsname

\title{The effect of substrate induced strain on the charge-ordering transition in
Nd$_{0.5}$Sr$_{0.5}$MnO$_{3}$ thin films}
\author{W. Prellier, Amlan Biswas, M. Rajeswari, T. Venkatesan and R. L. Greene}
\address{Center for Superconductivity Research, University of Maryland,
College Park, MD 20742}
\maketitle

\begin{abstract}
We report the synthesis and characterization of Nd$_{0.5}$Sr$_{0.5}$MnO$_{3}$
thin films grown by the Pulsed Laser Deposition technique on 100 
-oriented
LaAlO$_{3}$ substrates. X-ray diffraction (XRD) studies show that the films
are 101 
-oriented, with a strained and quasi-relaxed component, the latter
increasing with film thickness. We observe that transport properties are
strongly dependent on the thickness of the films. Variable temperature XRD
down to 100 K suggests that this is caused by substrate induced strain on
the films.
\end{abstract}
]


Hole-doped manganites, with the general formula RE$_{1-x}$A$_{x}$MnO$_{3}$
(RE=rare earth, A=alkaline earth) exhibit a rich phase diagram as a function
of the doping concentration $x$ ~\cite{jin,xiong,rao,ji}. For certain values
of $x$ and the average A-site cation radius ($<r_{A}>$), the metallic state
below $T_{c}$ becomes unstable and the material goes to an insulating state.
This is due to the real space ordering of the Mn$^{3+}$ and Mn$^{4+}$ ions
in different sublattices ~\cite{zirak,rao2,chen}. Such a charge ordering
(CO) transition is associated with large lattice distortions. This
phenomenon has been observed in Nd$_{0.5}$Sr$_{0.5}$MnO$_{3}$ ~\cite
{kuwahara} where $<r_{A}>=1.236$ \AA\ \cite{shannon}. However, most of the
work published up to now has been on single crystals or ceramic samples ~%
\cite{kuwahara,biswas,kawano}. One study on Nd$_{0.5}$Sr$_{0.5}$MnO$_{3}$
thin films has been reported ~\cite{wagner} in which no CO behavior was
observed. Epitaxial thin films have properties similar to those of single
crystals and are also important for potential device applications. We have
investigated growth of thin films of Nd$_{0.5}$Sr$_{0.5}$MnO$_{3}$ and their
structural and physical properties. The properties of these films are
significantly different from those observed in single crystals of Nd$_{0.5}$%
Sr$_{0.5}$MnO$_{3}$. We have proposed a model to explain these differences.

Thin films of Nd$_{0.5}$Sr$_{0.5}$MnO$_{3}$ were grown using the pulsed
laser deposition (PLD) technique. The target used had a nominal composition
of Nd$_{0.5}$Sr$_{0.5}$MnO$_{3}$. The substrates were [100] LaAlO$_{3}$
(LAO), which has a pseudocubic crystallographic structure with a=3.79 \AA .
The laser energy density on the target was about 1.5 J/cm$^{2}$, and the
deposition rate was 10 Hz. The LAO substrate was kept at a constant
temperature of 820 C during the deposition which was carried out at a
pressure of 400 mTorr of flowing oxygen. After deposition, the samples were
slowly cooled to room temperature at a pressure of 400 Torr of oxygen.
Further details of the target preparation and the deposition procedure are
given elsewhere ~\cite{ji}. The structural study was done, at room
temperature, by X-ray diffraction (XRD) using a Rigaku diffractometer. Low
temperature XRD experiments were performed with a Siemens kristalloflex
X-ray diffractometer. DC resistivity was measured by a four-probe method and
magnetization was measured using a Quantum Design MPMS SQUID magnetometer.
The composition analysis, by Rutherford Backscattering spectroscopy (RBS),
indicate a stochiometric composition within error limits.

Fig.1 shows the $\theta -2\theta $ scan of the films for different
thicknesses in the region 45$^{o}$-50$^{o}$. A peak is observed at 46.5$^{o}$
for all thicknesses which corresponds to an out-of-plane parameter of 1.955
\AA . Another diffraction peak appears gradually for thicknesses above 1000
\AA . This latter peak corresponds to a lattice parameter of 1.92 \AA ~ for
a 2000 \AA ~ film. These two peaks correspond to two phases which we will
call phases A and B (where A indicates the phase with the larger lattice
parameter). The structure of bulk Nd$_{0.5}$Sr$_{0.5}$MnO$_{3}$ is
orthorhombic ({\em Pnma}) with a=5.43153 \AA , b=7.63347 \AA ~ and c=5.47596
\AA ~\cite{kuwahara}. The out-of-plane parameter of the B-phase is equal to
the $d_{202}$ of bulk Nd$_{0.5}$Sr$_{0.5}$MnO$_{3}$. Since this phase, which
corresponds to the bulk value, appears for larger thicknesses, we assume
that it is the relaxed part of the film with less influence from the
substrate. The A-phase is always present, even in the thinnest film, which
strongly suggests that it represents the strained phase of the initial
layers of the film. The B-phase appears as the thickness is increased, which
implies that the film is relaxed after a critical thickness of about 500
\AA\ (see inset of Fig. 1). Moreover, the film is [101]-oriented, i.e. with
the [101] axis perpendicular to the substrate plane, which is similar to an
a-axis orientation according to the cubic perovskite cell. Note, that this
orientation has already been seen in (Pr$_{0.7}$Ca$_{0.3}$)$_{1-x}$Sr$_{x}$%
MnO$_{3}$ thin films grown by magnetron sputtering \cite{mercey}. Detailed
studies by Transmission Electron Microscopy are in progress to confirm these
results.

The resistivities of three films with different thicknesses are shown in
fig. 2. The 200 \AA ~ film is insulating at all temperatures. This layer is
formed due to non-uniform distribution of the strain over the film and
oxygen defects ~\cite{sun}. On increasing the thickness of the film, the
resistivity shows the metal-insulator transition ($T_{MI}$) near 200 K,
which is significantly lower than the $T_{MI}$ observed in the bulk
compound. This is an effect which has been observed in as-grown thin films
of other CMR materials and is attributed to the substrate induced strain on
\begin{figure}
\centerline{
\psfig{figure=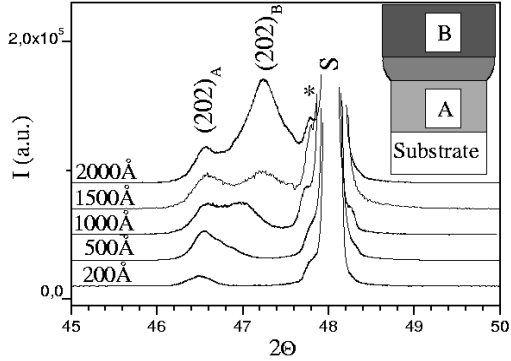,width=8.0cm,height=6.0cm,clip=}
}
\caption{Room temperature XRD of Nd$_{0.5}$Sr$_{0.5}$MnO$_3$ for different
thicknesses. The inset depicts phases A and B which are the strained and the
quasi-relaxed phases, respectively (see text). Peaks labeled * and S are due
to the sample holder and LaAlO$_{3}$ substrate.}
\end{figure}
\begin{figure}
\centerline{
\psfig{figure=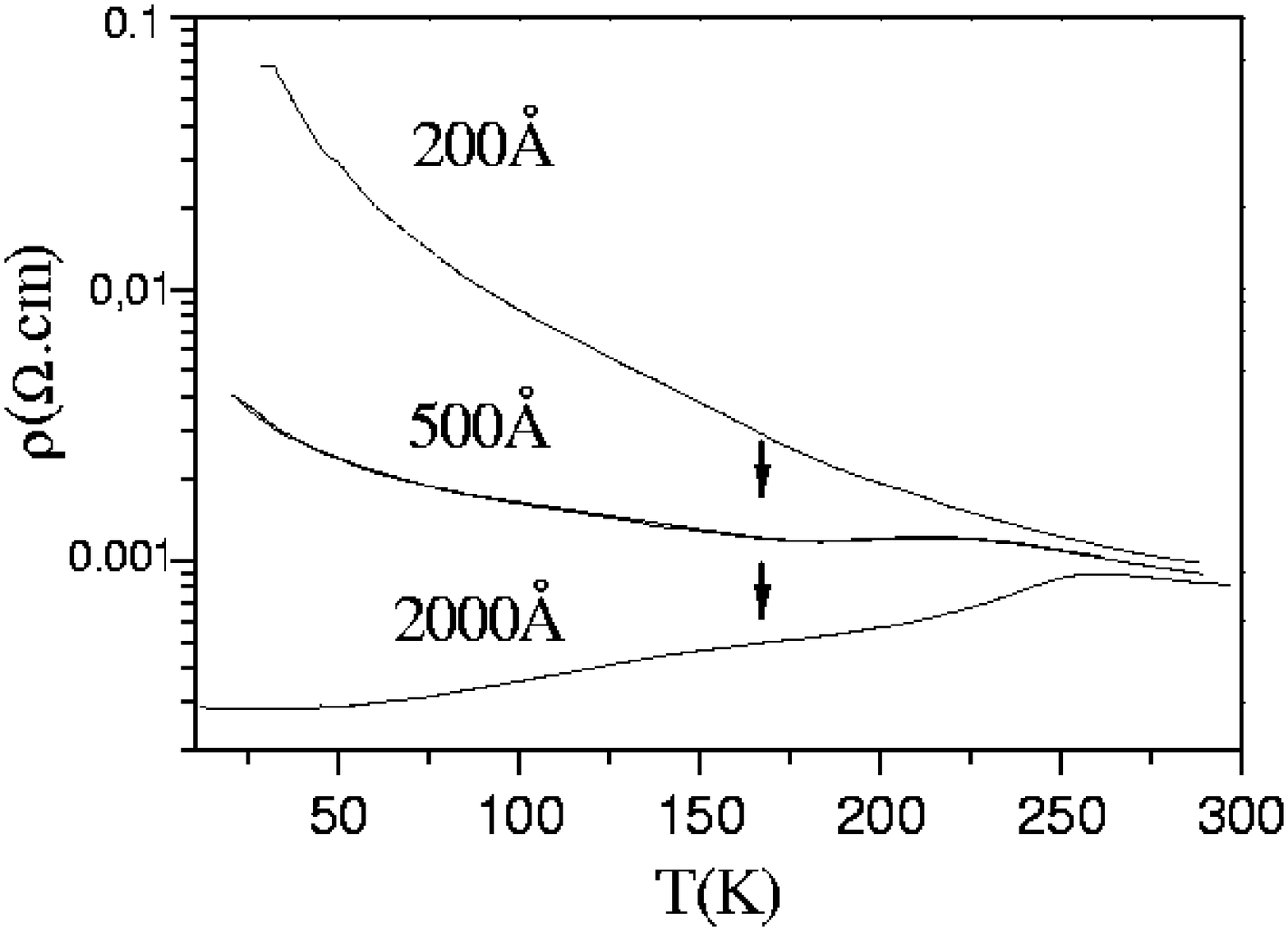,width=8.0cm,height=6.0cm,clip=}
}
\caption{Resistivity vs. temperature of films with different thicknesses.
Arrows indicate the T$_{CO}$ transition.}
\end{figure}
the film ~\cite{ji}. Also, around 170K, the resistivity starts increasing
with decrease in temperature. This is a typical signature of a charge
ordering transition ($T_{CO}$), but the rise in the resistivity is not as
sharp as observed in single crystals of Nd$_{0.5}$Sr$_{0.5}$MnO$_{3}$. The
effect of a magnetic field on the resistivity is shown in figure 3. A field
of 8 T shifts the $T_{CO}$ to 140 K and also shifts the $T_{MI}$ to higher
temperatures. There is a large reduction of the resistivity for the entire
temperature range. These features are qualitatively similar to the behavior
observed in single crystals of Nd$_{0.5}$Sr$_{0.5}$MnO$_{3}$. When the
thickness of the film is increased to 2000 \AA , the $T_{MI}$ shifts to a
higher temperature of 240 K as found in bulk. This is a signature of a
reduction of the strain on the film ~\cite{millis}. But now the CO
transition is suppressed and the resistivity shows a metallic behavior down
to the lowest temperatures. This is unexpected since, as the substrate
induced strain reduces, the properties should have approached those of the
bulk whereas we find film sample is semiconductor below T$_{CO}$.

The small deviation of composition between the film and the ideal
composition, Nd$_{0.5}$Sr$_{0.5}$MnO$_{3}$, is not enough to explain the
difference between the thin film and the bulk compounds. Two main features
need to be clarified :

(1) Why does the 500 \AA\ film show a CO-like behavior whereas the thicker
2000 \AA\ film is metallic at low temperatures?

(2) Why is the transition seen in the 500 \AA\ film not as sharp as seen in
bulk?

The thickness dependence of the properties suggests that strain plays an
important role in determining these properties. The substrate induced strain
can be expected to play an important role especially for the x=0.5
composition. In single crystals of Nd$_{0.5}$Sr$_{0.5}$MnO$_{3}$, there is
an abrupt change of the lattice parameters in the orthorhombic structure
which accompanies the CO transition ~\cite{kuwahara}. Is this abrupt change
seen in a thin film?

To answer these questions, we performed variable temperature XRD down to
100K in order to check the behavior of the out-of-plane parameter as the
temperature is lowered below the bulk $T_{CO}$. Fig.4 shows the evolution of
the $d_{202}$ between 110 K and 260 K for a 2000 \AA\ film. We also indicate
in this graph the evolution of the 202 reflection of a single crystal at low
temperatures, according to ref. \cite{kuwahara} (at room temperature the 202
reflection of the bulk coincides with that of phase B). From these data, we
can see that neither of the film phases has the bulk value of the lattice
parameters, below $T_{CO}$ of the bulk. Thus, it seems impossible to have a
sharp CO-transition in the films since the lattice parameters show no sharp
change at $T_{CO}$ nor do they have the same values as the bulk. In other
words, the lattice is constrained by the substrate and is unable to change
to the low temperature structure observed in single crystals.

The distortion of the lattice is essential for the CO transition, as this
decreases the Mn-O-Mn angle and thus the hopping probability, which results
in the insulating behavior. The strain does not allow the Mn-O-Mn angle to
change significantly in a thin film. However, the A-phase should exhibit a
resistivity behavior close to that of a single crystal at low temperatures
because the $d_{202}$ is close to the value of the single crystal below $%
T_{CO}$. It can 
be seen in fig. 2 that the 500 \AA\ film does show a
behavior resembling a charge-ordering transition, with the low temperature
resistivity similar to the resistivity behavior of a charge-ordering single
crystal under a hydrostatic pressure of about 1.5 GPa \cite{moritomo}. It is
known that the pressure induced by the substrate is of this order of
magnitude. Indeed, it is possible to stabilize in thin film form compounds
similar to those synthesized by high-pressure methods \cite{allen}. The 2000
\AA\ film is metallic
\begin{figure}
\centerline{
\psfig{figure=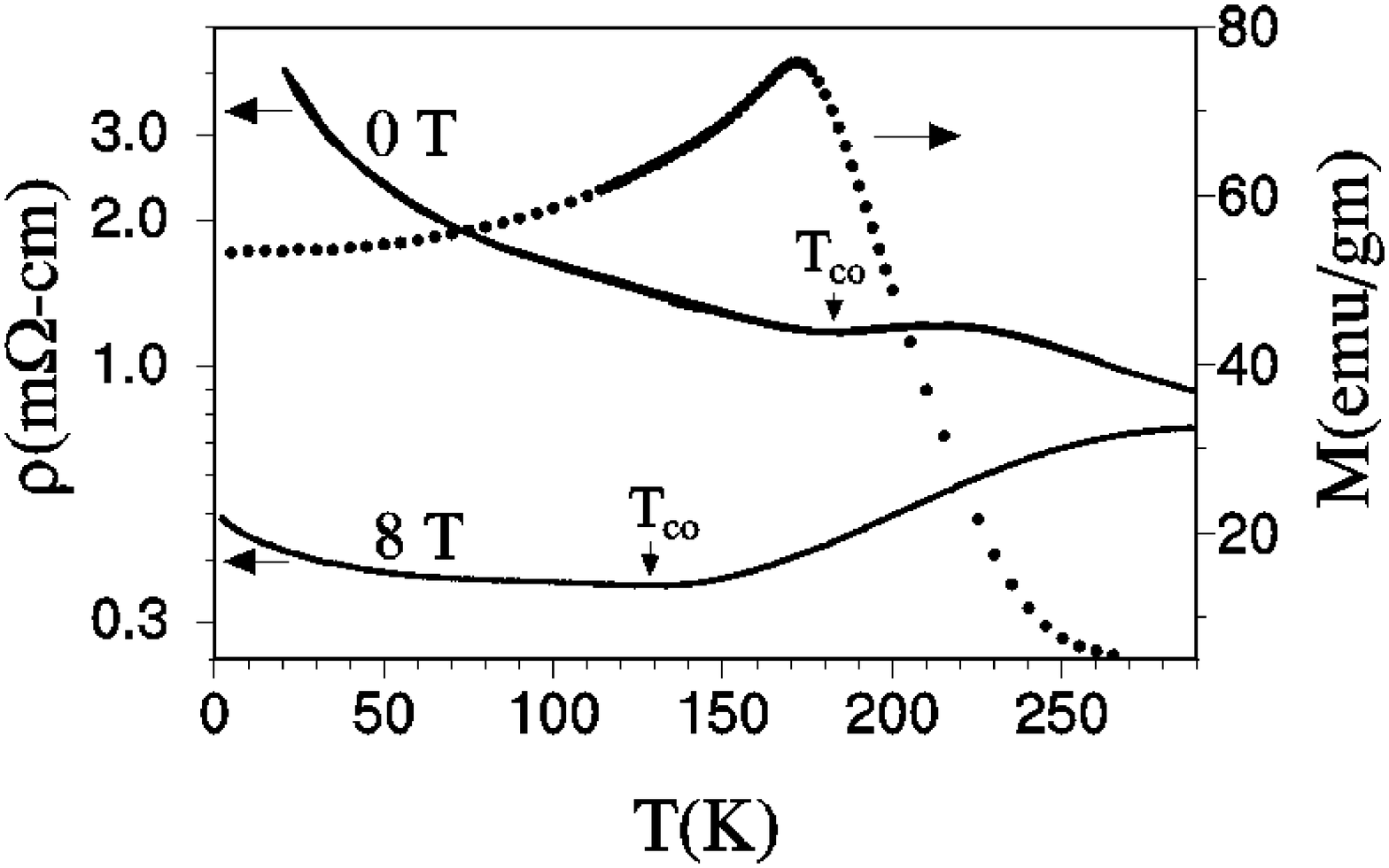,width=8.0cm,height=6.0cm,clip=}
}
\caption{Magnetic field dependence of the resistivity of the 500 \AA~ film
and the magnetization as a function of temperature measured in a field of
2000 Oe. }
\end{figure}
\begin{figure}
\centerline{
\psfig{figure=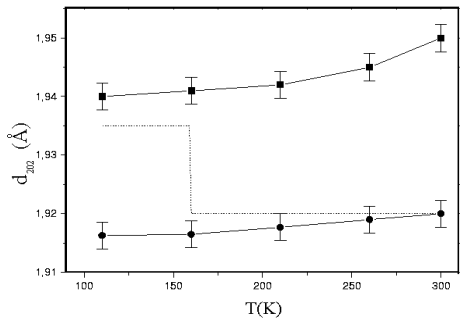,width=7.0cm,height=5.0cm,clip=}
}
\caption{Temperature dependence of d$_{202}$ for a 2000\AA~ film. Solid
squares and circles refer respectively to phases A and B (see text). Dotted
line corresponds to the parameter of a single crystal calculated from
Ref.[8].}
\end{figure}
because of the phase B which forms the top layer of
the thicker films. The phase B is the quasi-relaxed phase at room
temperature and has an out-of-plane parameter close to that of the bulk at
room temperature. This is reflected in the fact that the $T_{MI}$ of the
2000 \AA\ film is close to that of the single crystal. But when the
temperature is lowered the lattice parameters of the phase B do not change
as in a single crystal as seen (see Fig. 4). This keeps the film in a
metallic state. The overall resistivity behavior of the film is due to the
parallel combination of the phase A and phase B. When the thickness of phase
B exceeds a certain critical value, it changes the current distribution in
the film, effectively shorting out the underlying phase A. The effect of
phase A can still be seen in the resistivity curve of the 2000 \AA\ film, as
marked by the arrow.

In conclusion, we have grown Nd$_{0.5}$Sr$_{0.5}$MnO$_{3}$ thin films by PLD
on [100] LAO substrates. We found that the transport properties depend on
the thickness of the film. We show that for intermediate thickness, it is
possible to obtain a charge-ordering like behavior. However, thicker films
lead to a metallic behavior. This anomalous behavior is due substrate
induced strain on the film in the whole temperature range. Further
investigations are in progress, in particular, growing films on different
types of substrates, to confirm these results.

\acknowledgments
We acknowledge R. Ramesh and Y. Zheng for help in low temperature XRD
measurements. This work was partly supported by the MRSEC program of the NSF
(Grant \# DMR\ 96-32521).


\end{document}